\documentclass[aps,prc,twocolumn,superscriptaddress,nofootinbib,showpacs,showkeys,preprintnumbers]{revtex4}
\usepackage{graphicx}  
\usepackage{dcolumn}   
\usepackage{bm}        
\usepackage{amsmath, amsthm, amssymb}
\usepackage{lineno}
\usepackage{draftcopy}

\def\lsim{\mathrel{\rlap{
\lower4pt\hbox{\hskip-3pt$\sim$}}
\raise1pt\hbox{$<$}}}     
\def\gsim{\mathrel{\rlap{
\lower4pt\hbox{\hskip-3pt$\sim$}}
\raise1pt\hbox{$>$}}}     

\usepackage[bookmarks]{hyperref}

\begin{document}
\setlength{\linenumbersep}{5pt}

\title{Impact of the vector interaction on the phase structure of QCD matter}

\author{A. V. Friesen}
\affiliation{Joint Institute for Nuclear Research,  Dubna,
Russia}

\author{Yu. L. Kalinovsky}
\affiliation{Joint Institute for Nuclear Research,  Dubna,
Russia}

\author{V. D. Toneev}
\affiliation{Joint Institute for Nuclear Research,  Dubna,
Russia}

\begin{abstract}
The 2-flavor Polyakov-loop extended model
is generalized by taking into account the effective four-quark
vector-type interaction with the coupling strengths, which are
endowed with a dependence on the Polyakov field $\Phi$. The
effective vertex generates entanglement interaction between the
Polyakov loop and the chiral condensate. We investigate the
influence of an additional quark vector interaction and the
entanglement interaction on the location of the critical end-point
at the given chemical potential or quark density. It is shown that
the finite value of the vector interaction
strength $G_{\rm v}$ improves  the
model agreement with the lattice data. The influence of
the non-zero $G_{\rm v}$ and entanglement on the thermodynamic observables
 and the curvature of the crossover boundary in
the $T-\mu$ plane  is also examined.
\end{abstract}


\maketitle

\section{Introduction}
The QCD phase diagram and transitions between quark and hadron
phases are in the focus of recent investigations in both
theoretical and experimental fields of heavy energy physics.
However, investigation of the matter at finite temperature and real
chemical potential is difficult. Although the lattice theory
has made progress last years in calculations at the finite chemical
potential, the effective models are still most useful, for
example, Nambu-Jona-Lasinio-like models.

On the basis of the NJL model it is possible to build a phase
diagram describing the chiral restoration phase transition. There
are a crossover phase transition at low and intermediate $\mu$
and the first order transition at large $\mu$. The NJL model can
describe chiral symmetry breaking, but not the confinement
mechanism. The PNJL model is designed~\cite{Fu04} to make it
possible to treat both the mechanisms. To describe the quark-gluon
coupling and confinement transition, the  Polyakov loop has to be
included. The PNJL model can reproduce results of lattice QCD at
zero and imaginary $\mu$, where LQCD has no sign problem. However,
the PNJL and lattice results for calculation of chiral transition
temperature at $\mu = 0$ do not coincide. Then it was suggested 
renormalizing the lattice transition temperature for pure gluon
system $T_0 = 0.27$ GeV as $T_0 = 0.19$ GeV to get a realistic 
transition temperature  ($T_c^{\rm lat} \sim 0.16-0.19$ GeV)~\cite{PNJL_Ratti06}. 
It leads to a divergence of the chiral
transition and deconfinement transition temperatures at the zero
chemical potential in PNJL which do not coincide with the lattice results too.
To improve this situation, it was suggested that strong
correlations (entanglement) between the chiral and deconfinement
crossover transitions should be introduced by including an additional dependence of the
four-quark interaction on the Polyakov loop~\cite{GT_SakaiPRD82,
GT_SakaiPRD84}. The model was called later the entanglement-PNJL
(EPNJL) model.

In the PNJL model, the correlation between quark and gauge fields
is weak, so that the chiral and deconfinement crossover
transitions do not coincide without any fine-tuning of parameters
\cite{v_Sakai08}. For the zero chemical potential, the scalar type
eight-quark interaction is necessary to obtain a coincidence
between the two transitions,  or the vector-type four-quark
interaction is needed~\cite{v_Sakai08}. This fact indicates that a
true correlation between the quark condensate and the Polyakov
loop is stronger than that in the standard PNJL model appearing
through the covariant derivative between quark and gauge fields. The
problem of the existence of the critical end point and first order transition
on the phase diagram is still under consideration. The inclusion
of the vector interaction in the NJL or PNJL models was discussed, for
example in~\cite{v_Vogl91,v_Hatsuda94,Bu05}. In
Refs.~\cite{v_Kashiwa08,v_Sakai08}, it was shown that the equation
of state depends on the vector coupling constant $G_{\rm v}$. When
the vector interaction is taken into account, the first order
transition area becomes smaller, the critical point appears at
a higher chemical potential and lower temperature and is even
capable of removing the first order phase transition completely
from the phase diagram.

This paper studies the effects of the inclusion of the vector
quark interaction and its entanglement with the Polyakov loop,
their interplay as well as their influence on the thermodynamic
behavior of a quark system.


\section{PNJL with vector interaction}

We use the standard PNJL Lagrangian  with the vector interaction
term
\begin{eqnarray}
\label{Lpnjl} \mathcal{L}_{\rm
PNJL} &=& \bar{q}\left(i\gamma_{\mu}D^{\mu}-\hat{m}_0 \right) q +
G_s \left[\left(\bar{q}q\right)^2+\left(\bar{q}i\gamma_5
\vec{\tau} q \right)^2\right] \nonumber \\
 &-&G_{\rm v}(\bar{q}\gamma_\mu q)^2
-\mathcal{U}\left(\Phi[A],\bar\Phi[A];T\right)~.
\end{eqnarray}
A local chirally symmetric scalar-pseudoscalar four-point
interaction of quark fields $q,\bar{q}$ is introduced with an
effective coupling strength $G_s$, and the vector interaction
strength is regulated  by $G_{\rm v}$; $\vec{\tau}$ is the vector
of the Pauli matrices in flavor space, $\hat{m}_0$ is the diagonal
matrix of the 2-flavor current quark masses $m^0_u = m^0_d=m_0$.
The quark fields are coupled to the gauge field $A^\mu$ through
the covariant derivative
\begin{eqnarray}  \label{qPrel}
  D^\mu=\partial^\mu -iA^\mu.
   \end{eqnarray}
 The gauge coupling $g$ is
conveniently absorbed in the definition $A^\mu(x)=g{\cal
A}_a^\mu\frac{\lambda_a}{2}$ where ${\cal A}_a^\mu$ is the $SU(3)$
gauge field and $\lambda_a$ is the Gell-Mann matrices. The gauge
field is taken in the Polyakov gauge $A^\mu = \delta_0^\mu A^0 =
-i\delta_4^\mu A_4$. The field $\Phi$ is determined by the trace
of the Polyakov loop $L(\vec{x})$ and its conjugate
\begin{eqnarray}
\Phi[A] &=&\dfrac{1}{N_c} \mbox{Tr}_c L(\vec{x}) \nonumber\\
&=& \dfrac{1}{N_c}
\mbox{Tr}_c \left\lbrace \mathcal{P} \exp \left[ \displaystyle i
\int_{0}^{\beta} d \tau A_4 (\vec{x}, \tau) \right] \right\rbrace
\end{eqnarray}
with $\beta=1/T$ being the inverse temperature and $\mathcal{P}$
being the path ordering. In the absence of quarks, we have
$\Phi=\bar{\Phi}$ and the Polyakov loop servers as an order
parameter for deconfinement.  The Polyakov loop $\Phi$ is an exact
order parameter of spontaneous $Z_3$ symmetry breaking in pure
gauge theory. Although $Z_3$ symmetry is not exact in the
system with dynamical quarks, it still seems to be a good
indicator of the deconfinement phase transition. Therefore, we use
$\Phi$ to define the deconfinement phase transition.

An effective potential
$\mathcal{U}\left(\Phi[A],\bar\Phi[A];T\right)$ describes the
gauge sector and must satisfy the $Z(3)$ center symmetry;
therefore, one can choose any form satisfying the symmetry conditions
\cite{v_Dutra13}, but in this work the following general polynomial
form with the same parameters is used \cite{PNJL_Ratti06}:
\begin{eqnarray}\label{effpot}
\frac{\mathcal{U}\left(\Phi,\bar\Phi;T\right)}{T^4} &=& -\frac{b_2\left(T\right)}{2}\bar\Phi \Phi-
\frac{b_3}{6}\left(\Phi^3+ {\bar\Phi}^3\right)+
\frac{b_4}{4}\left(\bar\Phi \Phi\right)^2  \nonumber \\
b_2\left(T\right) = &a_0&+a_1\left(\frac{T_0}{T}\right)+a_2\left(\frac{T_0}{T}
\right)^2+a_3\left(\frac{T_0}{T}\right)^3. \label{Ueff}
\end{eqnarray}
$T_0=0.27$ GeV is the critical temperature of the deconfinement
transition in the pure gauge limit when quarks are assumed to be
infinitely heavy.

The grand  potential density for the PNJL ($N_f=$2) model in the
mean-field approximation  can be obtained from the Lagrangian
density (\ref{Lpnjl}) as \cite{PNJL_Ratti06,PNJL_Hansen07}:
\begin{eqnarray}
\hspace{-0.3cm}
\label{potpnjl} \Omega (\Phi, \bar{\Phi}, m, T, \mu) &=&
\mathcal{U}\left(\Phi,\bar\Phi;T\right) + G_s \sigma ^2 + G_{\rm
v} \rho^2 +\Omega_q,
\end{eqnarray}
where $\sigma =  \langle \bar{q} q \rangle $ is the quark condensate,
$\rho = <\bar{q} \gamma_0 q>$ is the quark density and the quark term is
\begin{eqnarray}
\Omega_q &=& -2 N_c N_f \int \dfrac{d^3p}{(2\pi)^3} E_p \nonumber \\
&-& 2N_f T \int \dfrac{d^3p}{(2\pi)^3} \left[ \ln N_\Phi^+(E_p)+
\ln N_\Phi^-(E_p) \right]~
\label{omegaq}
\end{eqnarray}
with the functions
\begin{eqnarray}
&& N^+_\Phi = \left[ 1+3\left( \Phi +\bar{\Phi} e^{-\beta
E_p^+}\right) e^{-\beta E_p^+} + e^{-3\beta E_p^+}
\right], \\
&& N^-_\Phi = \left[ 1+3\left( \bar{\Phi} + {\Phi} e^{-\beta
E_p^-}\right) e^{-\beta E_p^-} + e^{-3\beta E_p^-} \right]~,
\end{eqnarray}
where $E_p=\sqrt{{\bf p}^2+m^2}$ is the quasiparticle energy of
the quark; $E_p^\pm = E_p\mp \tilde{\mu}$ and
$\tilde{\mu}$ is related with the quark chemical potential $\mu$
and the quark density  through $G_{\rm v}$.

After performing the hadronization procedure the constituent quark
mass $m$ can be obtained as a solution of the gap equation (by
minimization of  $\Omega$ with respect to $m$: $\partial
\Omega/\partial m = 0$) . This condition is equivalent to the gap
equation~\cite{PNJL_Ratti06, PNJL_Hansen07}. There is also an
equation for $ \tilde{\mu}$ obtained as $\partial \Omega/\partial
{\rho} = 0$.
\begin{eqnarray}
\label{gap1}
m & = & m_0 - 2 G_s \ \sigma \label{masq}~,\\
\tilde{\mu} & = &\mu - 2 G_{\rm v} \rho.
\end{eqnarray}
Moreover, for PNJL calculations we should find values of
$\Phi$ and $\overline{\Phi}$ by minimizing $\Omega$ with respect
to $\Phi$ and $\overline{\Phi}$~\cite{PNJL_Hansen07} at given $T$
and $\mu$.

For the mass gap equation we get
\begin{eqnarray}
m = m_0 + 4 G_s N_c N_f \int_{\Lambda} \dfrac{d^3p}{(2\pi)^3}
\dfrac{m}{E_p} \left[ 1 - f^+ - f^- \right]
\end{eqnarray}
and for $\tilde{\mu}$
\begin{eqnarray}
\tilde{\mu} = \mu  - 4 G_{\rm v} N_c N_f \int_{\Lambda}
\dfrac{d^3p}{(2\pi)^3} \dfrac{m}{E_p} \left[ f^+ - f^- \right]
\end{eqnarray}
with the modified Fermi-Dirac distribution functions for fermions and antifermions
\begin{eqnarray}
f^+ &=& \left[\left( \Phi +2\bar{\Phi} e^{-\beta E_p^+}\right) e^{-\beta E_p^+}
+ e^{-3\beta E_p^+}\right]/ N_\Phi^+,
\label{fermimod}\\
f^- &=& \left[\left(\bar{\Phi}+2{\Phi} e^{-\beta E_p^-}\right) e^{-\beta E_p^-}
+ e^{-3\beta E_p^-} \right]/ N_\Phi^-.
\label{afermimod}
\end{eqnarray}
 One should note that if $\Phi \rightarrow 1$, Eqs.~(\ref{fermimod}),(\ref{afermimod})
 reduce to the standard NJL model.

\begin{figure} [h]
\centerline{
\includegraphics[scale=0.85]{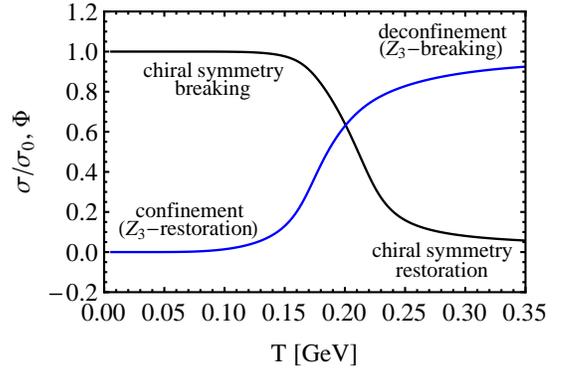}}
\caption{Condensate normalized to $\sigma_0=\sigma(T=0,\mu=0)$ and the Polyakov loop
as order parameters for the  chiral phase transitions and
confinement, respectively.}
 \label{condens_Phi}
\end{figure}

To obtain  free parameters in NJL-like models, the vacuum
values of $m_ \pi = 0.139$, $\sigma^{1/3} = -0.25$ GeV and $f_\pi
= 0.092$ should be reproduced. In this work, the parameters $\Lambda =
0.639$ GeV, $m_0 = 0.0055$ GeV and $G_s = 5.227$ \ GeV$^{-2}$  and
the same parameters for the effective potential  as in
\cite{PNJL_Ratti06, PNJL_Hansen07} are used. The parameter $T_0$ is
renormalized to $T_0 = 0.19$ GeV to make the critical temperature
lower (see \cite{PNJL_Ratti06, PNJL_Hansen07}).

From a theoretical point of view the QCD phase diagram on the ($\mu,
T$)-plane has the following structure: there are a hadron phase
of confined quarks and gluons and the quark-qluon phase where
deconfinement takes place. Both these phases are separated by a
crossover transition at low density or chemical potential and
temperature $\sim 0.17$ GeV. When the chemical potential reaches
a critical value, the crossover turns into the first order
transition. At the turning point named the critical end point the
second order transition takes place.

The $Z_3$-symmetry is explicitly broken in the presence of quarks with
the noninfinite mass and non-vanishing Polyakov loop. Really, the
phase structure of the quark models under discussion is determined
by the behavior of the order parameters $\sigma$, $\Phi$ and $\bar
\Phi$ as a function of temperature and quark chemical potential.
In the $\mu=0$ case it is illustrated in Fig.~\ref{condens_Phi}.
The Polyakov loop $\Phi$ vanishes at low temperature where the
confinement takes place. At the same time, the chiral symmetry is
explicitly broken by non-zero quark mass (at low temperatures).
Confinement implies spontaneous brake of the chiral symmetry. Are the
spontaneous brake of chiral symmetry imply confinement? Do they
connected? Lattice QCD computations with 2+1 flavors gave  the chiral transition temperature
 $T_{c}^{\rm lat}\sim 0.19$ GeV \cite{Tc_Cheng} (or 0.160 Gev in more recent
computations \cite{Tc_Borsanyi, Tc_Bazanov}). Pure gauge QCD on
the lattice without quarks gives the first order  deconfinement
transition at $T_0 = 0.27$ GeV. When the light quarks are added,
$Z_3$ symmetry is implicitly broken and deconfinement transition
takes place at $\sim 0.2$ GeV.

\begin{figure}[h]
\centerline{
\includegraphics[scale = 0.85] {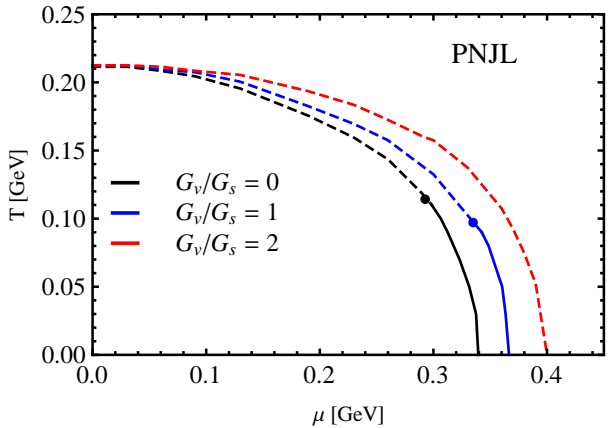}
} \caption{Phase diagram of the PNJL model with vector interaction
 for the fixed ratio $G_{\rm v}/G_s$.}
 \label{diagPNJL}
\end{figure}

As noted above, the order parameter of the spontaneous chiral
symmetry is the quark condensate $\sigma$. In the Wigner-Weyl
realization the condensate ''melts'' and disappears above a
characteristic  transition temperature. In the NJL and PNJL models the
crossover transition temperature is indicated as a maximum of
$\dfrac{\partial\sigma}{\partial T}$ for each $\mu$. The first
order phase transition is defined as ${\rm max}\dfrac{\partial
\rho}{\partial \mu}$ for each T.

The inclusion of the vector interaction in the NJL or PNJL models
discussed, for example in \cite{v_Vogl91,v_Hatsuda94,v_Kashiwa08,v_Sakai08},
leads to the dependence of the equation of state  on the
vector coupling constant $G_{\rm v}$.  The first order transition
area becomes smaller and the critical point appears at higher
chemical potential and at lower temperature. In
Fig.~\ref{diagPNJL},  the phase diagram of hadron matter is shown
for the PNJL model with vector interaction at different values of
$G_{\rm v}/G_s$. The case $G_{\rm v}/G_s=0$ displays a familiar
picture of the chiral crossover terminating as a second order
transition at a critical point. It was obtained that if the vector
coupling $G_{\rm v}/G_s> 1.5$, the first order transition region
disappears completely  and there exists a crossover transition
only.


\section{Entanglement of deconfinement and chiral symmetry}

One of the problems of the phase diagram construction is the so-called
pseudo-critical temperature $T_c$ definition at the low chemical
potential. Lattice calculations defined that $T_c^{\rm lat} =
0.173 \pm 8$ GeV; however $T_c$ from NJL calculation is higher.
Particularly, in the PNJL model $T_c$ depends on the parameter of the
effective potential $T_0 = 0.27$ GeV (see Eq.~(\ref{Ueff})). When
$T_0$ is renormalized to reach agreement with lattice results,
the chiral phase transition  and deconfinement transition  take place
at different temperatures, while the
\begin{figure} [h]
\centerline{
\includegraphics[scale = 0.85] {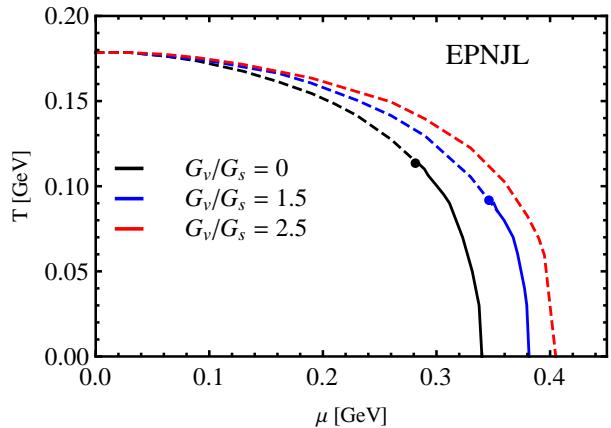}
} \caption{Phase diagram of the EPNJL model with vector
interaction for the fixed ratio $\tilde{G}_{\rm v}/\tilde{G}_s$.}
 \label{diagEPNJL}
\end{figure}
\begin{figure*}[ht]
\centerline{
\includegraphics[scale = 0.5] {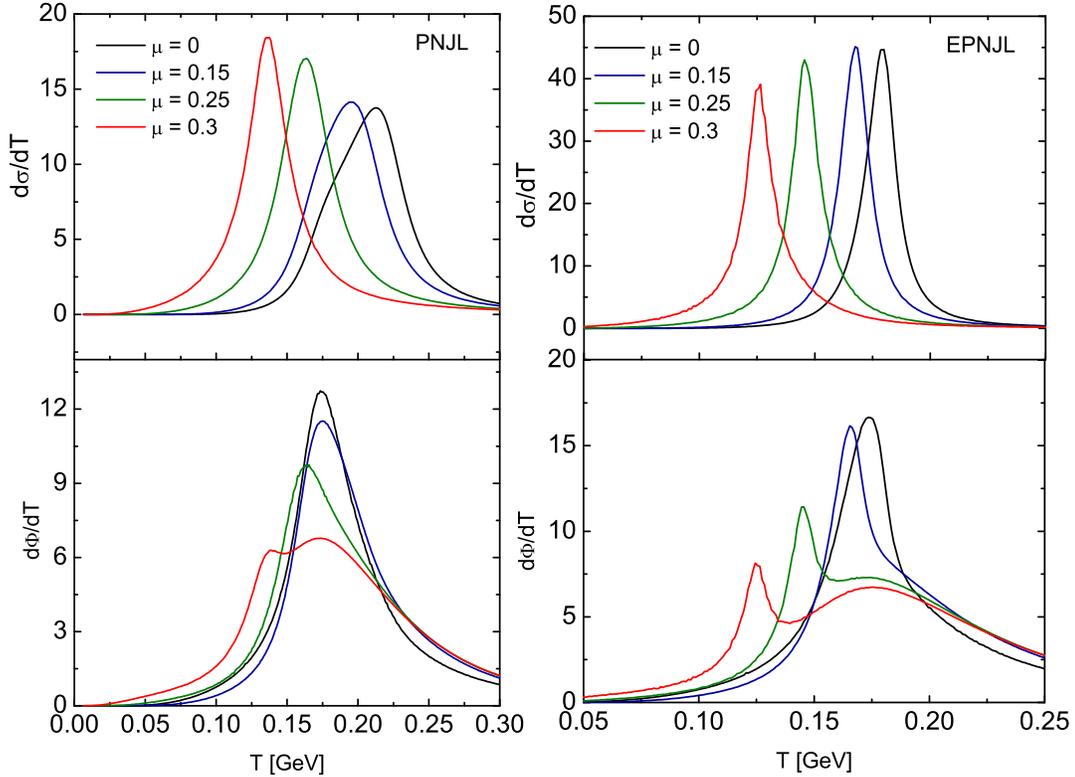}
} \caption{Temperature dependence of the order parameters for
various chemical potentials in the PNJL (left panel) and the EPNJL
model (right panel) with the vector interaction for
$\tilde{G}_{\rm v}/\tilde{G}_{\rm s} = 1.5$.} \label{diag-order}
\end{figure*}
lattice results show their coincidence. In \cite{PNJL_Ratti06,
PNJL_Hansen07}, it was suggested to define the pseudo-critical temperature as
$T_c=\dfrac{T_\sigma + T_\Phi}{2}$, where $T_\sigma = \dfrac{\partial\sigma}{\partial T}$ is the chiral transition temperature and $T_\Phi = \dfrac{\partial\Phi}{\partial T}$ is 
the deconfinement transition temperature. However, it does not lead
to the improvement of the situation with the difference between
the crossover and deconfinement
temperatures, because the entanglement between the chiral and
deconfinement transitions given only by Eq.~(\ref{qPrel}) is weak.
As it has been explained above, LQCD has the ''sign'' problem in
calculation at finite $\mu$. One of the approaches to solve the
problem is the introduction of the imaginary chemical potential
\cite{LQCD_Tc_Forcrand, LQCD_Elia1, LQCD_Elia2, LQCD_Chen,
LQCD_Wu}. Following the lattice imaginary potential in
\cite{v_Sakai08, Im_Sakai09} the imaginary chemical potential was
introduced for vector-type four-quark and eight-quark
interactions  in Refs.~\cite{v_Kashiwa08,v_Sakai08}. Later it
was suggested that there should exist entanglement
between chiral and deconfinement transition~\cite{GT_SakaiPRD82}.
So the renormalized scalar and vector four-quark couplings were
introduced as
\begin{equation}
\tilde{G_s}(\Phi) = G_s[1 - \alpha_1 \Phi\bar{\Phi} -\alpha_2(\Phi^3 + \bar{\Phi}^{3})],
\label{GsT}
\end{equation}
where the parameters $\alpha_1 = \alpha_2 = 0.2$ and $T_0 = 0.19$ GeV
are chosen to reproduce the  LQCD data for $\mu = 0$
\cite{LQCD_Tc_Forcrand}. In Refs.~\cite{GT_SakaiArXiv,GT_Costa},
the entanglement as Eq.~(\ref{GsT}) is used for a real chemical
potential. Following~\cite{GT_Sugano14}, the vector interaction
coupling constant can be introduced as
\begin{equation}
\tilde{G_{\rm v}}(\Phi) = G_{\rm v}[1 - \alpha_1 \Phi \bar{\Phi}
-\alpha_2(\Phi^3 + \bar{\Phi}^3)]~
\end{equation}
and therefore the scalar and vector couplings $\tilde{G}_{\rm v}$ and
$\tilde{G}_s$ become temperature-dependent in this model. To keep the ratio
$\tilde{G}_{\rm v}/\tilde{G}_s$ independent of
$\Phi$, we use the same $\alpha_1, \alpha_2$  as for $\tilde{G_s}$.
\begin{figure} [thb]
\centerline{
\includegraphics[scale = 0.85] {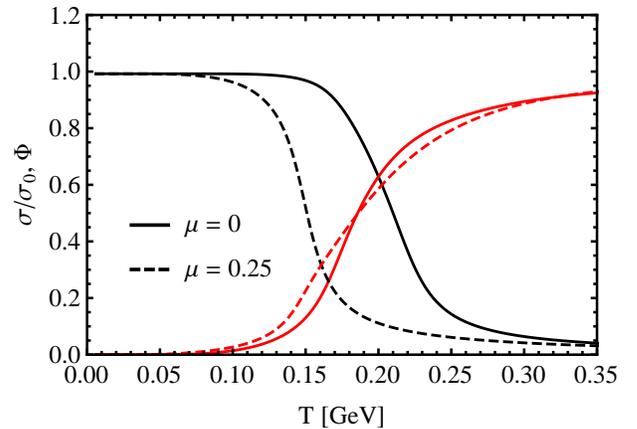}
} \caption{The normalized condensate and Polyakov loop for the PNJL
model at two different chemical potentials.} \label{two_peaks}
\end{figure}

The phase diagram for the EPNJL model is shown in
Fig.~\ref{diagEPNJL}. Similarly to the case of the PNJL model
(Fig.~\ref{diagPNJL}) the modified vector interaction splits the
phase curves $T-\mu$ and this splitting is the larger the higher
the vector coupling. In the case of the EPNJL model the critical
temperature at $\mu=$0 turns out to be the same as in the lattice
calculations, and the appropriate critical end-points are situated
at higher temperature than in the PNJL case.
\begin{figure*} [htb]
\centerline{
\includegraphics[width = 7.5cm] {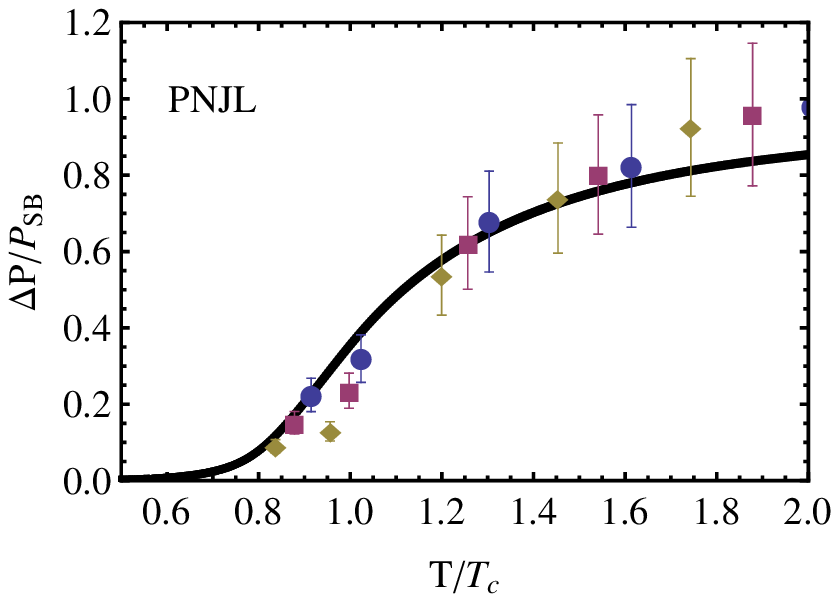}
\includegraphics[width = 7.8cm] {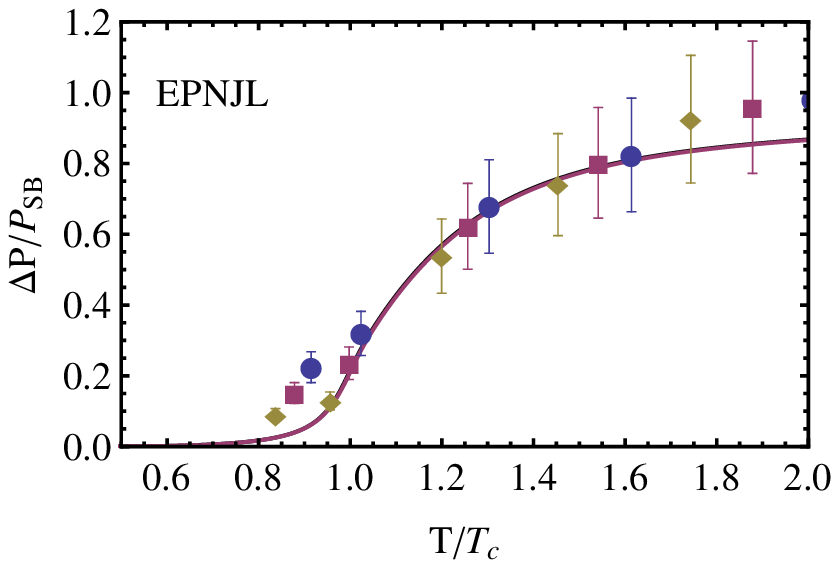}
} \caption{Comparison of the normalized pressure with the lattice
results for PNJL and EPNJL models at $\mu = 0$ GeV. In both models
the pressure is calculated for $G_v = 0$ and $G_v = 2 G_s$. In the
EPNJL case  both lines coincide. Lattice results are
from~\cite{KH01}}
 \label{m_latp}
\end{figure*}

The order parameters for both models are presented
in~Fig.~\ref{diag-order}.  It is clearly seen that  in the PNJL
model the locations of maxima defined as temperasture derivatives of $\sigma$
and $\Phi$ (compare the upper and bottom panels in the left column)
are very different at low $\mu$ and therefore  the critical chiral
and deconfinement temperatures are far from coincidence. The reason
for that is the renormalization of $T_0$. However, if entanglement
is taken into account, the peak positions coincide even at low
$\mu$ (see the right panel in~Fig.~\ref{diag-order})  Moreover, in
this case without entanglement one can observe from
Fig.~\ref{diag-order} (left bottom panel) that the derivative
$\partial \Phi/\partial T$ has two peaks at chemical
potential near the critical value. The existence of two peaks can
be related to coupling between the equations $\partial
\Omega/\partial m = 0$ and $\partial \Omega/\partial
\Phi = 0$ (see Fig.~\ref{two_peaks} for comparison). The first
peak in $\partial\Phi/\partial T$ is associated with the
deconfinement transition and comes from the abrupt fall in the
quark condensate $\sigma$. It is related to the Polyakov loop and
its temperature coincides with the chiral transition temperature. The
second peak arises from the Polyakov loop dynamics itself
\cite{v_Dutra13, Kahara08}.

 It can be seen that at the zero
chemical potential the areas of falling are closer to each other
(solid lines), but at the high chemical potential the falling of quark
condensate takes place earlier than for the Polyakov loop (dashed
lines), which leads to arising of two peaks.

\section{Comparison with lattice QCD and thermodynamics of models}

 The investigation of thermodynamics of a system starts
with the grand canonical ensemble, which is related to the
Hamiltonian $H$ as follows:
\begin{eqnarray}
 \label{can}
e^{-\beta V \Omega} = \mbox{Tr}\,\, e^{-\beta (H-\mu N)},
\end{eqnarray}
where $N$ is the particle number operator, $\mu$ is the quark
chemical potential and the operator $\mbox{Tr}$ is taken over
momenta as well as color, flavor and Dirac indices. The considered
models have the thermodynamic potentials defined from
Eq.~(\ref{potpnjl}). It has the vacuum part that does not vanish
as $T\rightarrow 0$ and $\mu \rightarrow 0$
\begin{eqnarray}
\label{omvac}
\Omega_{vac} = \frac{(m -m_{0})^2}{4G} - 2N_c N_f\int
\frac{d^3p}{(2\pi)^3}E_p.
\end{eqnarray}

 Basic thermodynamic quantities such as the pressure $P$,  energy
density $\varepsilon$,  entropy density $s$,  density of
quark number $\rho$ and the specific heat  $C_{\rm V}$ - can be
defined from Eq.(\ref{can}) as:
\begin{eqnarray} \label{p}
P &=& -\frac{\Omega}{V}, \\ \label{s} s &=& -\left(\frac{\partial
\Omega}{\partial T}\right)_\mu, \\ \label{epsilon} \varepsilon &=&
-P + Ts +\mu \, \rho, \\ \label{n} \rho &=& -\left(\frac{\partial
\Omega}{\partial \mu}\right)_T, \\ \label{Cv} C_V &=&
\frac{T}{V}\left(\frac{\partial s}{\partial T}\right)_V ~.
\end{eqnarray}

\begin{figure*} [ht]
\centerline{
\includegraphics[width = 7.5cm] {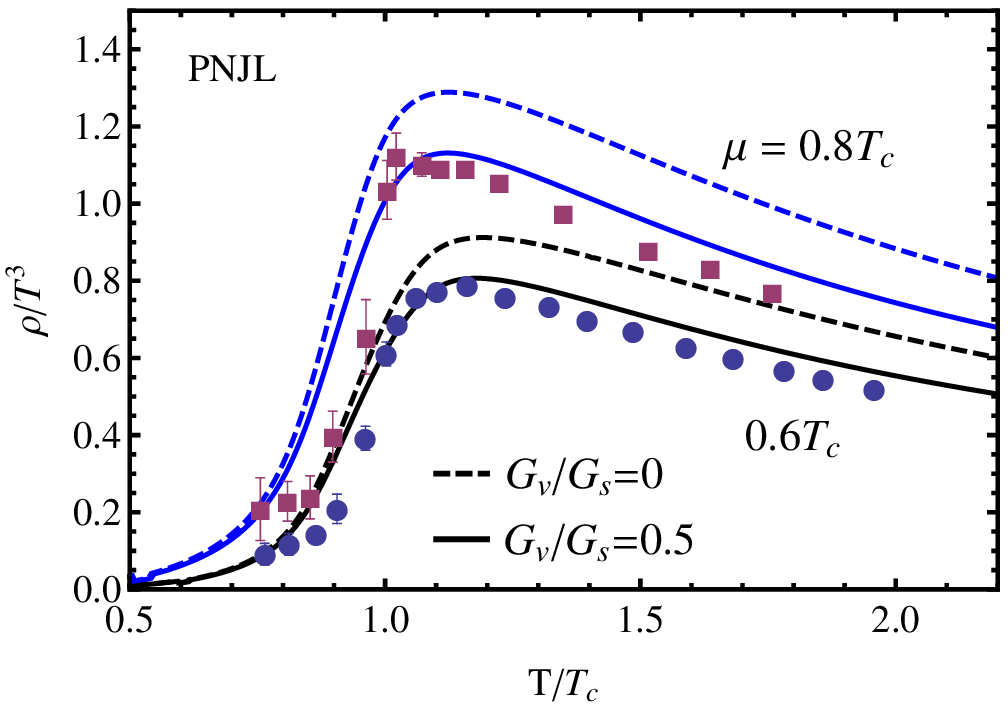}
\includegraphics[width = 7.5cm] {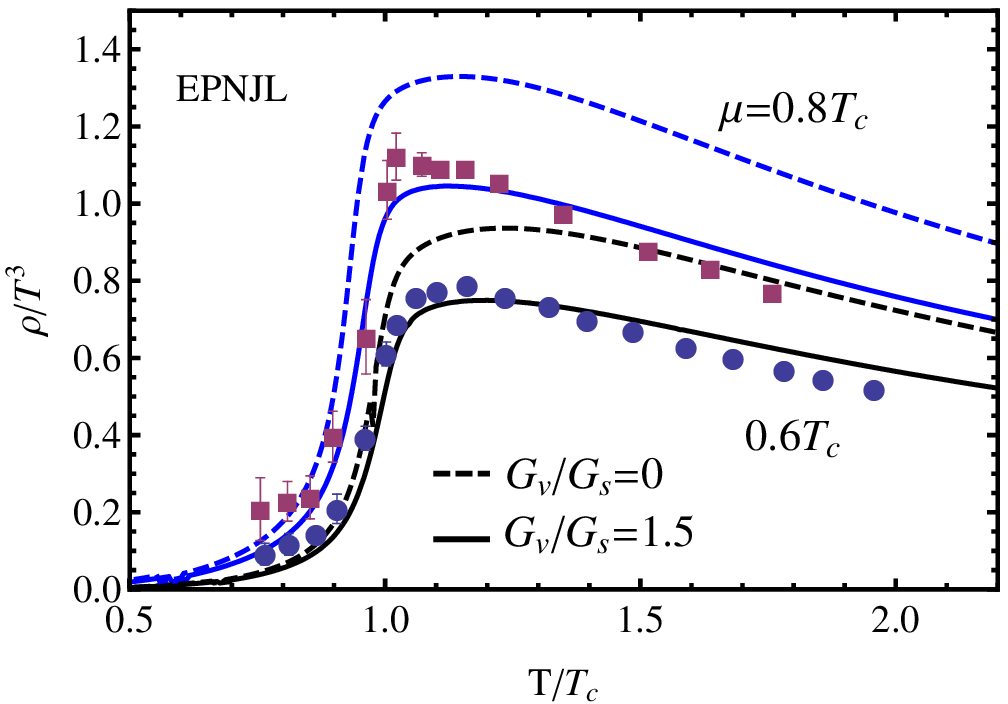}
} \caption{Comparison between the lattice results~\cite{Al03} and
the PNJL/EPNJL models for the normalized quark density at $\mu$ = 0.6
and 0.8 GeV.  }
 \label{m_latn}
\end{figure*}
\begin{figure} [h]
\centerline{
\includegraphics[scale = 0.9] {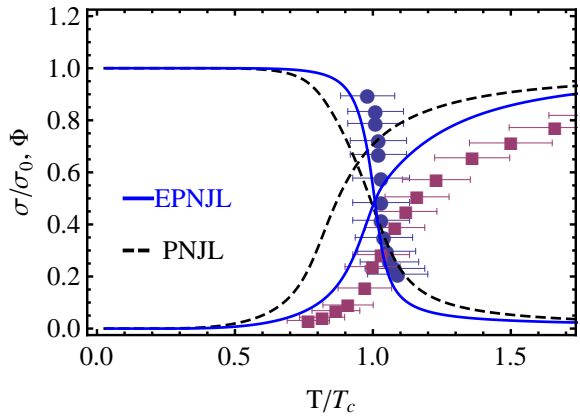}
} \caption{Comparison of the normalized chiral condensate and
Polyakov loop with the lattice results~\cite{KS04} at $\mu=$0 and
$G_{\rm v}=$0.}
 \label{m_lat}
\end{figure}

 In order to obtain the physical thermodynamic potential, which
corresponds to vanishing pressure and energy density at
$(T,\mu)=(0,0)$, one has to renormalize the thermodynamic
potential by subtracting its vacuum expression, Eq.~(\ref{omvac}).
This corresponds to the following definition of the physical
pressure:
\begin{eqnarray}
\frac{\Delta P}{T^4} = \frac{P(T, \mu, m) - P(0, 0, m)}{T^4}.
\end{eqnarray}

In Fig.~\ref{m_latp}, the reduced pressure for both models is shown as
a function of temperature at various $G_{\rm v}$.
The lattice results are taken from \cite{KS04,Mlat_Kaczmarek} for
the zero chemical potential. As is seen, at the zero chemical potential
 the results for $G_{\rm v} (\tilde{G_{\rm v}})
= 0$ and $G_{\rm v}(\tilde{G_{\rm v}})\neq 0$ just coincide since
there is no $n_q$ in Eq. \ref{p}. In
contrast, the normalized quark density is quite sensitive to the vector
coupling. The dependence of quark density on the chemical potential
$\mu$ and the vector coupling $G_{\rm v}$ is shown in Fig.~\ref{m_latn}.
The presented results clearly demonstrate that for agreement with
the LQCD the vector coupling $G_{\rm v}$ should be finite and grow
with $\mu$  reaching
approximately 0.5$G_s$ and 1.5$G_s$ in the PNJL and EPNJL models,
respectively. The strength of $G_{\rm v}$ was estimated in Ref.~\cite{Im_Sakai09}
from the 2-flavor LQCD results for the deconfinement transition line at the
imaginary chemical potential within the PNJL model including
scalar eight-quark interaction in  addition to the vector and scalar
four-quark interactions. They suggested that $G_{\rm v}/G_{\rm s}\approx$0.8
in this model. A similar analysis based on the non-local PNJL
model gives $G_{\rm v}/G_{\rm s}\approx$0.4 \cite{KHW11}. Recently,  it
was concluded within the 3-flavor PNJL model  that $G_{\rm v}$ is nearly
zero~\cite{SS14}. In the $G_{\rm v}$ estimates mentioned~\cite{Im_Sakai09,KHW11,SS14},
the entanglement effect is not considered.

The temperature behavior of the order parameters, quark condensate
$\sigma$ and the  Polyakov loop $\Phi$ is compared to the lattice data
in Fig.~\ref{m_lat}. Both models qualitatively reproduce the
lattice experiment, but the agreement with lattice is better for
the EPNJL description.

\begin{figure} [hb]
\centerline{
\includegraphics[scale = 0.78] {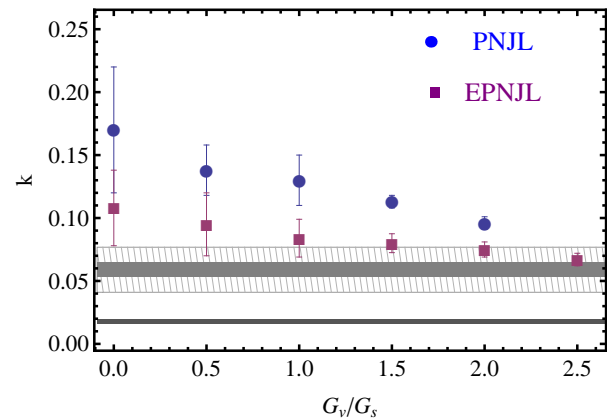}
} \caption{Curvature $k$ of the crossover line for the PNJL and EPNJL
models as a function of $G_{\rm v}/G_s$. The light-grey and grey bands
are the (2+1)-flavor lattice QCD results from~\cite{k_Fodor12} and
\cite{CCP14}, respectively. The large shaded band  corresponds to
calculations with the imaginary chemical potential~\cite{k_Karsch11}.}
 \label{k_koef}
\end{figure}

As has been noted above, in finite chemical potential calculations
there  exists  "sign" problem  in Lattice QCD.  However, the
introduction of the imaginary chemical potential allows one to solve the
problem. In Ref.~\cite{k_Fodor12}, it is suggested that the
critical end-point be searched as a common point where the phase transition
lines for all observables (the Polyakov loop, the strange quark
number susceptibility, the chiral condensate and the chiral
\begin{figure*} 
\centerline{
\includegraphics[width = 7.5cm] {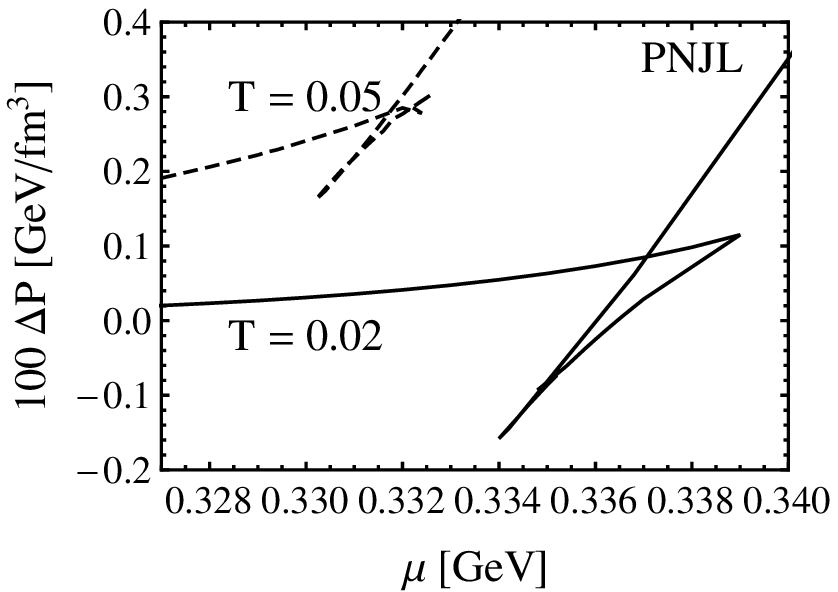}
\includegraphics[width = 7.5cm] {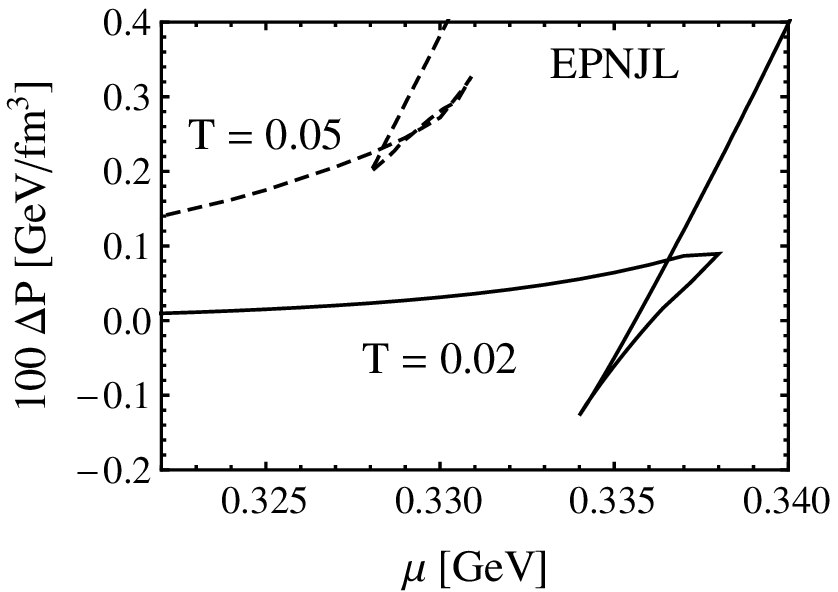}
} \caption{Pressure quark chemical potential dependence at
different temperatures in PNJL and EPNJL models. }
 \label{p-mu}
\end{figure*}
\begin{figure*} [thb]
\centerline{
\includegraphics[width = 7.50cm] {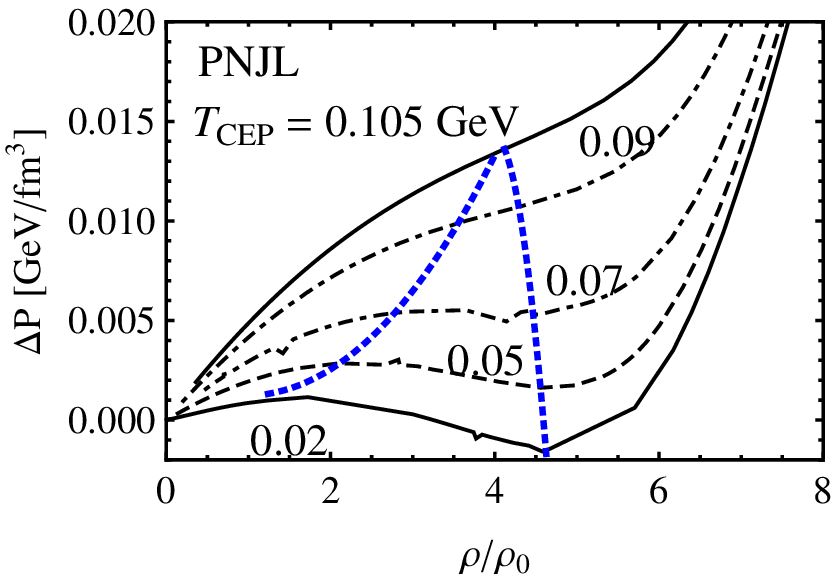}
\includegraphics[width = 7.50cm] {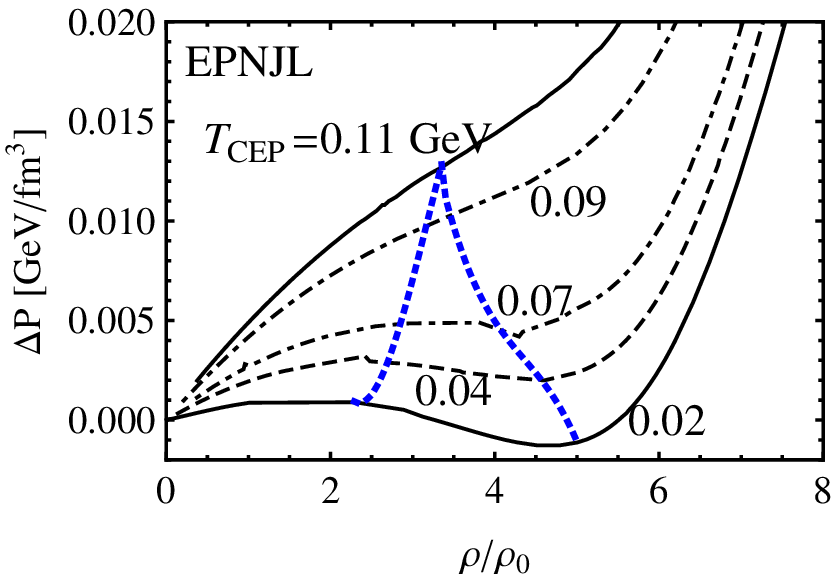}
} \caption{Pressure-quark density dependence at different
temperatures in the PNJL and EPNJL models. Dashed lines show the
unstable spinodal  boundaries. } \label{p-nq}
\end{figure*}
susceptibility) converge. The simplest parameter of
convergence is the curvature of the transition line
\begin{equation}
\frac{T_c(\mu)}{T_c(0)} = 1 - k \left(\frac{\mu}{T_c(\mu)}\right)^2.
\end{equation}
The curvature $k$ can be obtained in lattice QCD simulations in various ways
by suitable combinations of expectation values computed at $\mu=$0: by the
Taylor expansion method, analytical continuation for small and real $\mu_B$,
reweighting technique, determining the pseudo-critical line for purely
imaginary values of $\mu_B$ as
discussed for example in Ref.~\cite{Bo14}.
According to the (2+1)-flavor lattice results for Taylor expansion
technique~\cite{k_Karsch11}, $k=0.059\pm 0.020$ and also $k = 0.0089\pm0.0014$
and $k = 0.0066\pm 0.0020$ deduced from the scaling  properties of the chiral
condensate  and its susceptibility, respectively~\cite{k_Fodor12}.
 The (2+1)-flavor QCD results with the imaginary chemical potential $k=0.0132\pm 0.0018$
obtained from the renormalized chiral condensate and renormalized chiral
susceptibility $k=0.0132\pm0.003$~\cite{Bo14} seem to be compatible with
those cited above and obtained from Taylor coefficients.
In Fig.~\ref{k_koef}, the lattice measurements
of the crossover line curvature $k$ are plotted together with the
PNJL and EPNJL model results  as a function of $G_{\rm v}/G_{\rm s}$.
The noted above shifting of the chiral crossover to larger
$\mu$ implies a flattening of the curvature
of the crossover boundary in the neighborhood of $\mu=0$. Standard NJL and PNJL
calculations without vector interaction  usually produce a
curvature parameter $k$ that is too large in comparison with
the lattice results. Evidently, reasonably large vector coupling
strength  $G_{\rm v}/G_{\rm s}\gsim 1$ is capable of approaching such a small
curvature in the case of EPNJL but not for PNJL.

The equation of state of quark matter is displayed in
Figs.~\ref{p-mu}, \ref{p-nq}. The mixed phase as coexistence of
the hadronic and quark phases defined by the Maxwell construction
manifests itself as a straight line connecting two branches in the
triangle-loop formed near the critical point in the dependence of
pressure on the chemical quark potential
(Fig.~\ref{p-mu}). Two ends of this straight line correspond to
the boundary of the coexisting phase. Assuming here the
constant density within this range, one can define the isothermal
sound velocity,
$$\nu_T^2=\frac{\rho}{P_T+\epsilon} \left. \frac{\partial P}{\partial \rho}\right|_T.$$
These found numerically points (spinodals)
\begin{figure*} [thb]
\centerline{
\includegraphics[width = 7.5cm] {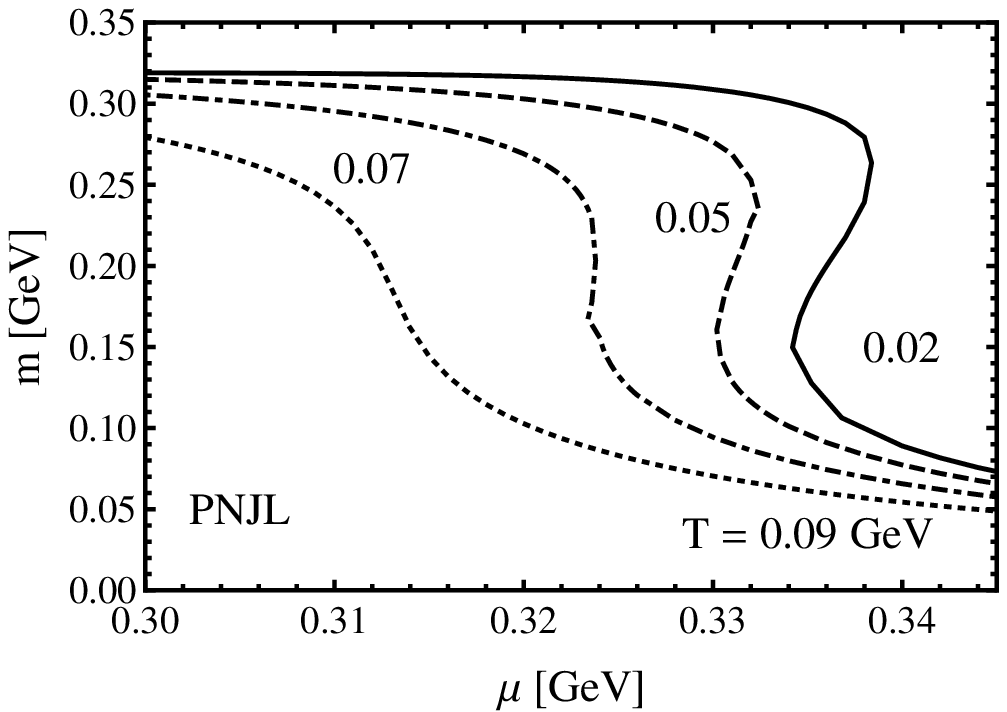}
\includegraphics[width = 7.5cm] {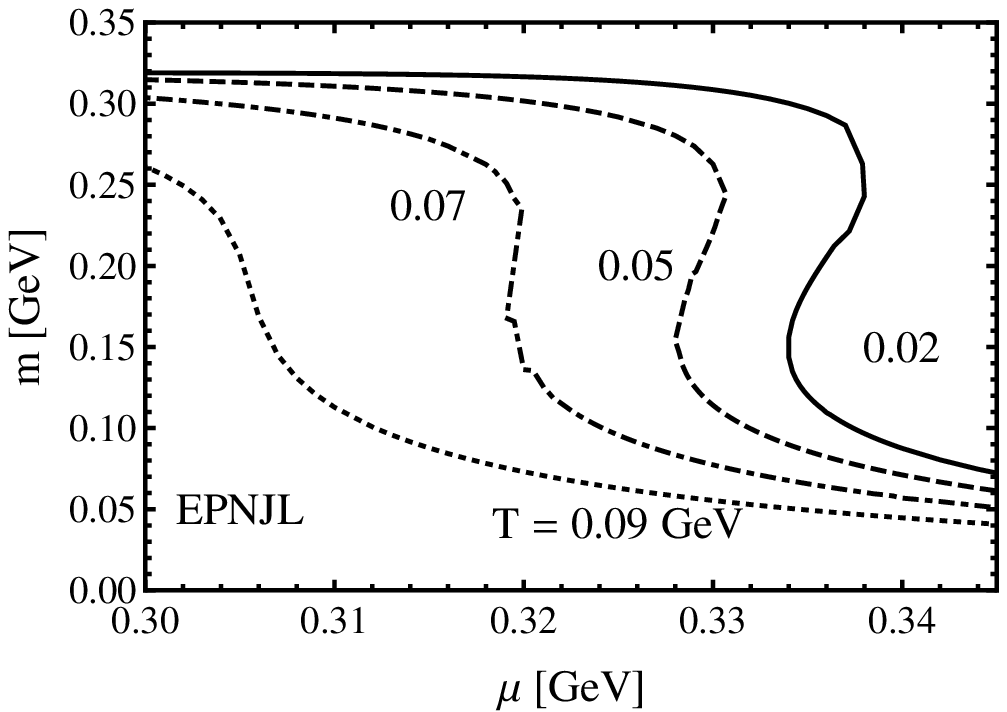}
} \caption{Quark mass vs the quark chemical potential at different
temperatures in PNJL and EPNJL models.  } \label{m-mu}
\end{figure*}
 limit a
thermodynamically unstable spinodal phase region, where spinodal
separation between confined and deconfined phases could occur.
One should note that the spinodal region shrinks increasing $G_{\rm v}$, i.e.
at the fixed temperature the first-order phase transition gets weakened
and eventually becomes the second order. This explains why the critical
point moves to lower temperature and finally disappears from the phase
diagram.

 The spinodal region is shown in Fig.~\ref{p-nq} by the dotted line as
 a function of the reduced density. One can define the
boundaries of the unstable region also by the isoentropic sound velocity
$$\nu_s^2=\frac{\rho}{P_T+\epsilon} \left. \frac{\partial P}{\partial \rho}\right|_s$$,
but the high-temperature part of this region is smaller than that
for isothermal spinodality and does not reach the critical end
point~\cite{Ra05}.

As is seen from the comparison of Fig.~\ref{m-mu} with Fig.~\ref{p-nq},
this instability range strongly correlates with the range where
the quark mass derivative with respect to the chemical potential
becomes negative, $\partial m/\partial\mu<0$. Note that the
location of the critical point ($\sim$ 100 MeV) is lower than that
in lattice calculations.


\section{Concluding remarks}

 The 2-flavor Polyakov-loop extended model
is generalized by taking into account the effective four-quark
vector interaction and its dependence on the Polyakov loop. The
effective vertex generates entanglement interaction between the
Polyakov loop and the chiral condensate. We investigated the impact
of the entanglement interaction on thermodynamic properties of the
system and, in particular, on the location of the critical end point
at the given chemical potential or quark density. It is shown that
these effects improve the model agreement with the lattice data.

The inclusion of the vector four-quark interaction shifts the location
of the critical point towards lower temperature and larger
chemical potential as compared to the standard NJL models. Additional account for
entanglement between the deconfinement transition and the chiral symmetry restoration
moves the critical end point in the opposite direction but it is not enough for the
model to agree with the lattice result. While the exact value of the vector
coupling strength is uncertain, our results clearly point to a strong
vector repulsion in the hadronic phase (large $\mu$) and near-zero
repulsion in the deconfined phase ($\mu \to$0) supporting conclusions
of Ref.~\cite{SS14}. Our comparison between the lattice data and the
PNJL/EPNJL results for the curvature of the
crossover boundary supposes large values of $G_{\rm v}$.
At present, the vector coupling $G_{\rm v}$ cannot be determined from
experiment and/or lattice QCD simulations but, eventually, the combination
of neutron star observations and the energy scan of the phase-transition
signals in the large $\mu$ range at FAIR/NICA may provide
us with some hints on its precise
numerical value. While many authors consider $G_{\rm v}$ as a free parameter,
whose values range in $0.5\leq G_{\rm v}/G_{\rm s}\leq 0.5$ \cite{RSSV10,CNB10},
other try to fix it in different ways as in Refs. \cite{v_Dutra13,Im_Sakai09,KHW11,BHW13}
predicting $0.3\leq G_{\rm v}/G_{\rm s}\leq 3.2$, so that the true value
remains undetermined.

It is possible that some new interaction should
be added. In particular, the chiral phase transition  in an
extended Nambu-Jona-Lasinio model with scalar type eight-quark
interactions is studied in~\cite{KK07}. The scalar type nonlinear
term hastens the restoration of chiral symmetry, while the
scalar-vector mixing term makes the transition sharper. The scalar
type nonlinear term shifts the critical end point toward the values
predicted by lattice QCD simulations and the QCD-like
theory~\cite{KK07,v_Kashiwa08}. Note that a sufficiently strong
vector interaction is likely to provide the necessary repulsion at
high density that helps supporting two-solar-mass neutron
stars~\cite{OBG10}.

Our preliminary study shows that the discussed spinodal separation
between the confined and deconfined phases in fact occurs in
relativistic nuclear collisions and a more refined analysis is
needed.  To find signals of this spinodal separation is a
challenge for future work.

{\bf Acknowledgments} \\
 We are thankful to P. Costa  and M. Ilgenfritz for useful discussions.
This work has been supported in part by the LOEWE
center HIC for FAIR as well as by BMBF and by the RFBR, grant No.
13-01-00060a (Yu.L.K).  V.D.T.\ is also partly
supported  by the Russian Ministry of Science and
Education,  research project identifier
RFMEFI61614X0023 and by the Heisenberg-Landau grant of JINR.

\end{document}